\documentclass[sigplan, screen, 10pt, nonacm]{acmart}

\AtBeginDocument{%
  }

\renewcommand\footnotetextcopyrightpermission[1]{}
\usepackage{xspace}
\usepackage{minted}
\usepackage{pgfplots}
\usetikzlibrary{arrows.meta}
\usepackage{subcaption}
\usepackage[font={small,it}]{caption}
\usepackage{xspace}
\usepackage{cleveref}
\usepackage{svg}
\usepackage{adjustbox}
\usepackage{array}
\newcolumntype{R}[2]{%
    >{\adjustbox{angle=#1,lap=\width-(#2)}\bgroup}%
    l%
    <{\egroup}%
}
\usepackage{threeparttable}
\usepackage[most]{tcolorbox}
\usetikzlibrary{patterns}
\usepackage{soul}
\usepackage[normalem]{ulem}
\usepackage{tabularx}

\usepackage[subtle]{savetrees}
\usepackage{pifont}
\newcommand{\rot}[1]{\rotatebox{90}{\textbf{#1}}}
\newcommand{\cmark}{\ding{51}}%
\newcommand{\xmark}{\ding{55}}%

\newcommand{\allnotes}[1]{}
\renewcommand{\allnotes}[1]{#1}

\newcommand{\sys}{{\scshape Rex}\xspace}
\newcommand{\bench}{\sys}
\newcommand{\bNum}{{87}\xspace}
\newcommand{\numRealworldNotebook}{{12}\xspace}
\newcommand{\rNum}{{38}\xspace}
\newcommand{\Observable}{{\scshape Observable}\xspace}
\newcommand{\Marimo}{{\scshape Marimo}\xspace}
\newcommand{\Ipyflow}{{\scshape IPyflow}\xspace}
\newcommand{\RRAll}{{\scshape Rerun-All}\xspace}
\newcommand{\EC}{{\scshape Run Subsequent Cells}\xspace}
\newcommand{\ECGraph}{{\scshape Run Sub.}\xspace}
\newcommand{\ECGraphShort}{{\scshape Run Sub.}\xspace}
\newcommand{\IpyflowES}{{\scshape IPyflow}\xspace}
\newcommand{\RRAllFS}{{\scshape Rerun-All}\xspace}

\definecolor{MyBlue}{HTML}{0073b2}
\definecolor{MyLightBlue}{HTML}{56b4e9}
\definecolor{MyYellow}{HTML}{f0ee42}
\definecolor{MyGreen}{HTML}{009e73}
\definecolor{MyLightPurple}{HTML}{aa67ff}
\definecolor{MyBeige}{HTML}{cdb1ad}
\definecolor{MyDarkBeige}{HTML}{6C2E26}
\definecolor{MyDarkBlue}{HTML}{5d6e9e}
\definecolor{MyPurple}{HTML}{e2e6f2}
\definecolor{MyLightGreen}{HTML}{addd8e}
\definecolor{MyYellowHighlight}{HTML}{fbf7c5}

\sethlcolor{MyYellowHighlight} 

\DeclareRobustCommand{\tableCirc}{\tikz\draw[fill=black] (0,0) circle (2.5pt);\xspace}
\DeclareRobustCommand{\tableCircLegend}{\tikz\draw[fill=black] (0,0) circle (1.5pt);\xspace}
\pgfplotsset{my legend/.style={
    legend image code/.code={
        \filldraw [fill=#1, draw=black] (-0.05cm, -0.1cm)
        rectangle (0.2cm,0.2cm);
    },
}}

\pgfplotsset{legendblack/.style={
    legend image code/.code={
        \filldraw [fill=black, draw=black] (-0.05cm, -0.1cm)
        rectangle (0.2cm,0.2cm);
    },
}}

\pgfplotsset{legendnortheastlines/.style={
    legend image code/.code={
        \filldraw [pattern=north east lines, draw=black] (-0.05cm, -0.1cm)
        rectangle (0.2cm,0.2cm);
    },
}}

\usepackage[dvipsnames]{xcolor}
\usepackage{listings}
\usepackage{minted}
\tcbuselibrary{listings}

\newcommand{\heading}[1]{\vspace{2pt}\noindent\textbf{\emph{#1}}.\enspace}
\newcommand{\eg}{{\em e.g.}, }
\begin{document}
\fancyhead{}
\pagestyle{empty}

\title{When Do Reactive Notebooks Fail to React?}

\author{Yuchen Lu}
\authornote{Equal contribution}
\affiliation{ 
      \institution{Brown University}
      \country{}
    }
    \email{yuchen_lu@brown.edu}
\author{Megan Zheng}
\authornotemark[1]
\affiliation{ 
      \institution{Brown University}
      \country{}
    }
    \email{megan_zheng@brown.edu}

\author{Will Crichton}
\affiliation{ 
      \institution{Brown University}
       \country{}
    }
    \email{will_crichton@brown.edu}

\author{Akshay Narayan}
\affiliation{ 
      \institution{Brown University}
       \country{}
    }
    \email{akshayn@brown.edu}

\author{Deepti Raghavan}
\affiliation{ 
      \institution{Brown University}
       \country{}
    }
    \email{deeptir@brown.edu}

\author{Nikos Vasilakis}
\affiliation{ 
      \institution{Brown University}
    \country{}
    }
    \email{nikos@vasilak.is}



\begin{abstract}
Computational notebooks are convenient for programmers, but can easily become confusing and inconsistent due to the ability to incrementally edit a program that is running. 
Recent reactive notebook systems, such as \Ipyflow, \Marimo and \Observable, strive to keep notebook state in sync with the current cell code by re-executing a minimal set of cells upon modification.
However, each system defines reactivity differently.
Additionally, within any definition, we find simple notebook modifications that can break each system.
Overall, these inconsistencies make it difficult for users to construct a mental model of their reactive notebook's implementation.
This paper proposes \bench, a fine-grained test suite to discuss and assess reactivity capabilities within reactive notebook systems.
We evaluate \bench on three existing reactive notebook systems and classify their failures with the aims of (i) helping programmers understand when reactivity fails and (ii) helping notebook implementations improve.
\end{abstract}

\maketitle
\section{Introduction}
\label{introduction}

Computational notebooks are the de facto platform for interactive data exploration and data science: in 2020 alone, GitHub hosted over 10 million public notebooks~\cite{jetbrainsnotebooks}.
Notebooks enable users to incrementally edit a running program, modifying a program's state without restarting it from scratch. 
However, incremental edits are a double-edged sword; 
they can lead to confusing program states~\cite{chattopadhyay2020notebookpain, 9127201, 10.1145/3173574.3173606, grus2018dont, 10.1145/3173574.3173748}. 
To address this problem, some recent systems have added a notion of \emph{reactivity} where the notebook's runtime syncs its state  with user edits by automatically re-executing a subset of cells in the notebook~\cite{observable2025observable, marimo, ipyflow, pluto, vizierdb, livebook, HexNotebooks, guzharina2021revamped}. 

Reactive notebook systems advertise strong claims about their reactivity.
For example, \Marimo~\cite{marimo} claims that it ``guarantees your notebook code, outputs, and program state are consistent'';  \Observable~\cite{observable2025observable} and \Ipyflow~\cite{ipyflow} make similar claims.
A user may therefore reasonably expect that after editing a notebook cell, these systems will 
then rerun the appropriate set of cells needed to maintain consistency.

In actuality, these guarantees only extend to certain kinds of code based on how the system tracks dependencies between cells.
\Marimo and \Observable's guarantees only apply to programs which do not mutate data defined in earlier cells.
\Ipyflow allows cross-cell mutation but sometimes fails to react correctly due to limitations of its dependency analysis.
For all such systems, effects on the external environment (e.g., reads and writes to files) are out of scope.



Unfortunately, users of reactive notebooks may nonetheless write code which is either out of scope or falls into a soundness hole.
For example, \Cref{fig:intro-fig} shows how an innocuous edit can lead to an inconsistent state, which occurs in \Marimo, \Observable, and \Ipyflow at the time of writing.
These systems lack a mechanism for identifying such problematic code.
\Marimo and \Observable detect binding a variable twice, 
but not mutations to a data structure as in \Cref{fig:intro-fig}.
We argue that effective use of reactive notebooks should start with a clear understanding of these systems' limitations, leading to our question: \textbf{when are reactive notebooks not reactive?}

%

\input{graphs/intro_fig_new}
To answer this question, we introduce \sys (\S\ref{s:3}), a comprehensive benchmark suite that contains \rNum modifications to \numRealworldNotebook real-world notebooks
and \bNum synthetic micro-benchmarks.
Each benchmark contains a notebook, an edit, and an expected result under our definition of reactivity (\S\ref{s:2}), implemented in both Python and JavaScript where applicable.

We evaluate \bench on three state-of-the-art reactive notebook systems---\Marimo, \Observable, and \Ipyflow---against standard baselines like \RRAll and \EC (\S\ref{s:realworld}, \S\ref{s:synthetic}).
Our evaluation uncovers a trade-off between language flexibility and reactive reliability.
While dataflow-driven systems (\Marimo and \Observable) maintain perfect precision for direct assignments, they fail to capture or prohibit mutation-heavy code paths, allowing notebooks to enter inconsistent states. 
On the other hand, while \Ipyflow handles a wider range of mutations, it suffers from severe over-reactivity, over-executing unrelated cells by an average of 2.8$\times$ in realworld workloads and causing developers to wait unnecessarily for $\sim$25 minutes for independent pipelines to re-run.
Additionally, all evaluated systems remain blind to external environment and file system alterations.

\section{Defining Reactivity}
\label{s:2}
Given a modification to a notebook, we define a \emph{reaction} as the actions needed to make the notebook's state and environment \emph{consistent} with its current source code.
Runtimes typically implement reactivity by re-running a subset of the cells after a modification.
Specifically, the notebook's state and environment is \emph{consistent} with the current source code if it is identical to the program state and environment achieved by executing the notebook from scratch in a fresh environment.
However, consistency hinges on what exactly ``program state and environment'' encompasses.

Each of a notebook's cells can contain a combination of variable definitions, re-definitions, mutations, and read/write interactions with external state such as the file system or a database.
After each cell executes, a specific version of the program's state exists, including all memory in its heap, stack, and data segments.
While most developers write notebooks in a single language (\eg Python), programs often contain variables with pointers to memory managed and allocated by external native libraries (\eg pointers to each column in a Pandas dataframe).
In addition to memory, when  each cell runs, a specific version of the notebook's external environment exists, \eg the state of files the notebook reads/writes to or the state of external API services it interacts with.

The execution order of a notebook also affects reactivity.
Computational notebook runtimes like Jupyter typically allow running notebook cells in any order, multiple times.
Establishing a baseline state of restarting and rerunning the entire notebook from scratch requires defining some execution order.
Systems like \Marimo and \Observable actually implement reactivity without depending on a top-down, linear, execution order.
For the scope of this paper, we focus on reactivity within a linear, top-to-bottom execution order, where each cell is run once.
%

\subsection{Scope of Reactivity}
Based on the behaviors observed in real notebooks in \bench, we outline a scope for reactivity, in terms of the modifications allowed and programs allowed.

\heading{Programs and Modifications}
The ideal reactive system handles any arbitrary code in the original notebook as well as any modification to this notebook.
A modification to a notebook can add, modify, or delete a cell, or swap the position of two cells.
Modifications can change assignments or re-assignment of variables, change which specific state is being mutated, how state is being mutated, or any combination of these.
However, some notebooks explicitly define certain program structures as out-of-scope.
\Marimo and \Observable can handle notebooks with \emph{intra-cell mutation} (mutating a variable defined in the same cell), but cannot handle notebooks with \emph{inter-cell mutations} (mutating a variable defined in a previous cell).

\heading{Program State}
Ideally, reactive runtimes should achieve consistency for \emph{the entire internal program memory and external environment which the notebook interacts with}.
In terms of program memory, it is only important to achieve consistency for \emph{observed memory}.
When notebook cells are re-run or restarted from scratch, specific pointer addresses may change, but the program state visible to the developer will likely be the same.
Unfortunately, achieving consistency for the external environment is nearly impossible, as it may change independently of the notebook itself.
For example, the state of a file on the filesystem could change due to external edits in between the original notebook execution and the modification.

\bench covers notebooks where the program state consists of observed memory (including memory managed in native libraries) and external filesystem state that the notebook program exclusively interacts with.
We defer studying notebooks which interact with state that changes via out-of-band mechanisms, such as shared external API services, shared files, or the network, to future work.
Finally, in \bench, we assume notebooks use random seeds whenever relying on random number generators.

\subsection{Measuring Reactivity: Soundness and Precision}
A reaction is \emph{sound} if it produces a consistent state, and a reaction is \emph{precise} if no redundant work is performed to ensure consistency.
The bottom-right notebook in \Cref{fig:intro-fig} represents an inconsistent state: due to the stale mutation to \verb|x['a']|, a clean linear execution of the modified code could not produce the same state.
Runtimes that produce inconsistent states are not sound.
In contrast, a simple ``restart and run all'' approach would be sound but not precise, because re-executing the cell \verb|z = 123| is unnecessary.
An \emph{ideal reaction} is both sound and fully precise, and the goal of a reactive notebook system is to guarantee soundness while maximizing precision.

\heading{Undefined behavior}
Consistency is only meaningful for programs that fall within the scope of a given reactive system. 
For a given system, if certain program structures are out-of-scope (\eg cross-cell mutations), then a notebook containing those structures exhibits \emph{undefined behavior}.
For instance, the code pattern in \Cref{fig:intro-fig} triggers undefined behavior in \Marimo and \Observable, while it triggers defined (but unintended) behavior for \Ipyflow.
In \bench, we analyze reactivity for each system according to what it defines as in and out-of-scope.
For programs or modifications that are out-of-scope, we measure whether the notebook runtime warns the user about the undefined behavior.
In these cases, a sound reaction involves successfully warning the user about out-of-scope behavior.

\section{A Benchmark for Reactive Notebooks} 
\label{s:3}

To evaluate the soundness and precision of reactive notebook systems, we present \sys, a benchmark containing \rNum modifications across \numRealworldNotebook realworld machine learning notebooks and \bNum synthetic \texttt{ipynb} programs. 
Each entry provides a notebook, a specific modification, and the exact set of cells required to satisfy a sound and precise reaction.

Different modifications pose vastly different levels of complexity for reactive systems to handle correctly.
To reflect this, \bench categorizes modifications using a strict hierarchy, assigning each modification the highest-complexity operation it triggers, starting from the most straightforward.

Table~\ref{table:services} breaks down the modifications evaluated in \bench, for both the real-world and synthetic benchmarks.
\textbf{Direct assignment} modifies the definition of a variable on either the LHS or RHS of the assignment operation.
\textbf{Reassignment} redefines a variable first defined in another cell.
\textbf{Mutation} performs any in-place modification of existing values or external  resources, including files, or calling random-number generators.
Mutation is the hardest category: the variable binding remains unchanged even though the underlying state has been modified, making staleness invisible to symbol-level dependency tracking.
We further classify mutations into three subcategories by mechanism: 
external state, native mutations, 
and Python built-in mutations (detailed in Appendix \ref{app:mutation-categorization}).



\renewcommand{\topfraction}{0.9}	
\renewcommand{\textfraction}{0.1}	

\begin{table}[t] 
\center
\setlength\tabcolsep{5pt}
\caption{\textbf{Declared Reactivity Scope:} Overview of in-scope and out-of-scope for reactive notebook systems. \RRAll, \EC, and \Ipyflow are Jupyter-based Python notebook platforms; \Observable and \Marimo are proprietary platforms supporting JavaScript and Python respectively.}
\begin{threeparttable}
\begin{tabular}{lccccccc}
\toprule
\footnotesize
\textbf{Feature}
& \rot{\footnotesize \RRAll} & \rot{\footnotesize \ECGraph} & \rot{\footnotesize \Ipyflow} & \rot{\footnotesize \Marimo} & \rot{\footnotesize \Observable} &
\rot{\footnotesize Realworld Count} &
\rot{\footnotesize Synthetic Count} \\

\midrule
\addlinespace
Direct Assignment  & \cmark & \cmark & \cmark & \cmark & \cmark & 12 & 18 \\
Reassignment & \cmark & \cmark & \cmark & \tableCirc & \tableCirc & 1 & 14 \\ 
Mutation & \cmark & \cmark & \cmark & \xmark & \xmark & 25 & 55 \\
\hspace{5pt} (File System) & \cmark & \xmark & \xmark & \xmark & \footnotesize NA & 3 & 8 \\
\hspace{5pt} (RNG) & \cmark & \xmark & \xmark & \xmark & \footnotesize NA & 2 & 2 \\
\bottomrule
\end{tabular}
\begin{tablenotes}[flushleft]
    \footnotesize \item[(\cmark)] in-scope for reactive system 
    \footnotesize \item[(\tableCircLegend)] out-of-scope, users may not execute notebook with feature
    \footnotesize \item[(\xmark)] out-of-scope, but users may still execute notebook with feature

\vspace{-5pt}
\end{tablenotes}
\end{threeparttable}
\label{table:services}
\end{table}

\newcommand*\halfcirc[1][1ex]{%
  \begin{tikzpicture}
  \draw[fill] (0,0)-- (90:#1) arc (90:270:#1) -- cycle ;
  \draw (0,0) circle (#1);
  \end{tikzpicture}}

\newcommand*\circled[1]{\tikz[baseline=(char.base)]{
            \node[shape=circle,draw,inner sep=0.4pt] (char) {#1};}}

\heading{Realworld benchmarks} 
%
We reviewed approximately 100 notebooks from recent machine learning conference artifact repositories (\eg ICML and NeurIPS), excluding those with irresolvable dependency conflicts, missing local data, or GPU requirements, yielding \numRealworldNotebook fully reproducible notebooks.
%
For each, we used Claude 4.6 Sonnet to propose modifications targeting cross-cell dependencies, then manually reviewed and corrected each modification (prompts detailed in Appendix~\ref{app:prompt}), yielding \rNum distinct modifications.
%
To establish ground truth for precision evaluation, we manually reasoned through each notebook's execution topology to determine the minimal expected rerun set, then empirically validated each set by confirming the resulting runtime state matched a complete top-to-bottom execution from a fresh environment.
Both benchmarks include cells unrelated to the target modification to evaluate unnecessary re-execution. 


\heading{Synthetic micro-benchmark}
\bench includes \bNum synthetic micro-benchmarks to isolate specific language semantics across all three complexity categories, as the realworld benchmarks do not cover all corner cases.
%
We translated 52 benchmarks to JavaScript for \Observable evaluation; the remaining 35 used incompatible Python libraries or required file systems access unavailable in \Observable's purely in-browser runtime.
Of the 52 translatable benchmarks, 16 were incomparable due to \Observable's out-of-order execution~\cite{observable2021how} diverging from our definition of reactivity.
\Marimo benchmarks were produced using its \texttt{converter} tool~\cite{marimo2025comingfromjupyter}.
Table~\ref{table:services} summarizes the modification counts and declared scopes for all evaluated systems.

\subsection{Evaluation Methodology}
\label{s:eval-methodology}

To ensure a fair and consistent evaluation across all systems and both benchmark datasets, we utilize a standardized execution and scoring protocol.

\heading{Baseline Systems}
For baseline comparisons, we implemented two language-agnostic systems as JupyterLab extensions~\cite{granger2017jupyterlab}: \RRAll and \EC.
%
\RRAll clears all notebook state and re-executes cells top-to-bottom, modeling the common user action of rerunning all cells to sync state~\cite{chattopadhyay2020notebookpain}. 
It also resets the filesystem between modifications by deleting all notebook-created files, simulating systems aware of external resources.
%
\EC re-executes all subsequent cells located spatially below the modified cell.
It assumes that definitions always appear above their uses in the notebook, making cell position a sufficient approximate for dependency.
This makes \EC language-agnostic and a naive optimization over \RRAll. 

\heading{Benchmark systems}
We evaluate \Ipyflow and both baselines on both the real-world benchmark and the synthetic benchmarks.
\Observable operates in a JavaScript runtime incompatible with Python ML libraries such as PyTorch and NumPy, so we only evaluate \Observable within a subset of the synthetic benchmarks which we could translate.
\Marimo requires a fundamental rewrite of notebooks into a pure-functional dataflow paradigm forbidding cross-cell mutations, precluding evaluation on some of the real-world benchmarks, so we only evaluate \Marimo on the synthetic benchmarks.



\heading{Execution Protocol}
To evaluate each benchmark, we initiate two fresh kernels: one for the \textit{original notebook} and one for the \textit{modified notebook}. 
We execute the original notebook sequentially, apply the single target modification, and trigger the system's reactivity to measure which cells the system re-runs.
Then, we execute the modified notebook top-to-bottom and compare each cell's output to the same cell in the original notebook, to verify whether the re-run achieved the same output.
We collect the number of cells the reactivity system re-executes to measure precision.

\heading{Scoring Soundness}
As Table~\ref{table:services} shows, some systems declare only certain program structures as in-scope for reactivity and others further provide linting to prohibit execution when given out-of-scope features by raising an error. To fairly assess a system's efficacy with respect to its explicitly supported features, we consider \textit{out-of-scope but caught} as sound and behaviorally correct. In other words, when a system identifies and prohibits (errors on) any execution of out-of-scope features declared in its documentation, we consider the reaction as sound.
For all notebooks except \RRAll, we consider mutations caused by filesystem interactions and randomness as out-of-scope.
\Marimo and \Observable only allow direct assignment modifications, but catch cases when developers write reassignments.

\section{Real World Notebooks}
\label{s:realworld}

Evaluation with real world notebooks helps understand which common Python features are most important to support.

\begin{figure}[t!]
\centering
\begin{tabular}{c}

\ref{sharedlegendRW}
\\[-0.3em]
\hspace{-10pt}
\begin{tikzpicture}
    [
      /pgfplots/every axis/.style={
        title = (A) Direct Assignment (12),
        title style={at={(0.3,0)},anchor=north,yshift=-29pt, font=\footnotesize},
        height = 3cm,
        width = 3.3cm,
        symbolic x coords={\RRAllFS, \ECGraphShort, \IpyflowES},
        ytick={0, 5, 10, 12},
        legend to name=sharedlegendRW,
        legend style={text=black, xshift=5pt, anchor=center, legend columns=2, font=\footnotesize, nodes={anchor=west}},
        x tick label style={
            anchor=north,
            yshift=-8pt,
            xshift=-13pt,
            rotate=45,
            font=\tiny
        },
        xtick=data,
        ybar stacked,
        ymin=0,
        ymax=12,
        ylabel = {Count},
        ylabel style={yshift=-16, xshift=-5, font=\footnotesize},
        bar width=9pt,
        enlarge x limits= 0.2,
        ytick distance=5,
        yticklabel style = {font=\tiny},
        tick style={draw=none},
        axis x line*=bottom,
        axis y line*=left
    },
    ]
    \begin{axis}
        \addplot [fill=MyDarkBlue, draw=black, forget plot] coordinates {
            (\RRAllFS, 12) (\ECGraphShort, 11) (\IpyflowES, 12)
        };
        \addplot+[fill=MyDarkBeige, draw=black, forget plot] coordinates {
            (\RRAllFS, 0) (\ECGraphShort, 1) (\IpyflowES, 0)
        };
        \addplot+[fill=MyBeige, draw=black, forget plot] coordinates {
            (\RRAllFS, 0) (\ECGraphShort, 0) (\IpyflowES, 0)
        };
    \addlegendimage{my legend = MyDarkBlue}
    \addlegendentry{In-Scope Match}
    \addlegendimage{my legend = MyDarkBeige}
    \addlegendentry{In-Scope Mismatch}
    \addlegendimage{my legend = MyPurple}
    \addlegendentry{Out-of-Scope Caught}
    \addlegendimage{my legend = MyBeige}
    \addlegendentry{Out-of-Scope Not Caught}
    \addlegendimage{legendnortheastlines}
    \addlegendentry{Out-of-Scope Correct (RNG + Files)}
    \end{axis}
\end{tikzpicture}
\begin{tikzpicture}
    [
    /pgfplots/every axis/.style={
        title = (B) Reassignment (1),
        title style={at={(0.35,0)},anchor=north,yshift=-29pt, font=\footnotesize},
        height = 3cm,
        width = 3.3cm,
        symbolic x coords={\RRAllFS, \ECGraphShort, \IpyflowES},
        ytick={0, 1},
        x tick label style={
            anchor=north,
            yshift=-8pt,
            xshift=-13pt,
            rotate=45,
            font=\tiny,
        },
        xtick=data,
        ybar stacked,
        ymin=0,
        ymax=1,
        bar width=9pt,
        enlarge x limits= 0.2,
        ytick distance=1,
        yticklabel style = {font=\tiny},
        tick style={draw=none},
        axis x line*=bottom,
        axis y line*=left
    },
    ]
    \begin{axis}
        \addplot [fill=MyDarkBlue, draw=black, forget plot] coordinates {
            (\RRAllFS, 1) (\ECGraphShort, 1) (\IpyflowES, 1)
        };
        \addplot+[fill=MyDarkBeige, draw=black, forget plot] coordinates {
            (\RRAllFS, 0) (\ECGraphShort, 0) (\IpyflowES, 0)
        };
        \addplot+[fill=MyBeige, draw=black, forget plot] coordinates {
            (\RRAllFS, 0) (\ECGraphShort, 0) (\IpyflowES, 0)
        };
    \end{axis}
\end{tikzpicture}
\begin{tikzpicture}
   [
    /pgfplots/every axis/.style={
        title = (C) Mutation (25),
        title style={at={(0.35,0)},anchor=north,yshift=-29pt, font=\footnotesize},
        height = 3cm,
        width = 3.3cm,
        symbolic x coords={\RRAllFS, \ECGraphShort, \IpyflowES},
        ytick={0, 5, 10, 15, 20, 25},
        x tick label style={
            anchor=north,
            yshift=-8pt,
            xshift=-13pt,
            rotate=45,
            font=\tiny,
        },
        xtick=data,
        ybar stacked,
        ymin=0,
        ymax=25,
        bar width=9pt,
        enlarge x limits= 0.2,
        ytick distance=5,
        yticklabel style = {font=\tiny},
        tick style={draw=none},
        axis x line*=bottom,
        axis y line*=left
    },
    ]
    \begin{axis}
        \addplot [fill=MyDarkBlue, draw=black, forget plot] coordinates {
            (\RRAllFS, 25) (\ECGraphShort, 17) (\IpyflowES, 16)
        };
        \addplot+[fill=MyDarkBeige, draw=black, forget plot] coordinates {
            (\RRAllFS, 0) (\ECGraphShort, 3) (\IpyflowES, 4)
        };
        \addplot+[fill=MyBeige, draw=black, forget plot] coordinates {
            (\RRAllFS, 0) (\ECGraphShort, 1) (\IpyflowES, 4)
        };
        \addplot+[fill=MyDarkBlue,pattern=north east lines, draw=black, forget plot] coordinates {
            (\RRAllFS, 0) (\ECGraphShort, 4) (\IpyflowES, 1)
        };
    \end{axis}
\end{tikzpicture}

\end{tabular}
\caption{\textbf{Real world Correctness Results By Modification,} 
%
partitioned by whether systems correctly react to in-scope benchmarks (\textit{in-scope match} vs. \textit{in-scope mismatch}), or detect and alert users of out-of-scope features prior to execution to reduce undefined behaviors from reactivity (\textit{out-of-scope caught} vs. \textit{out-of-scope not caught}).}
\label{graph:by_modifications_realworld}
\vspace{-10pt}
\end{figure}

\subsection{Soundness}
\label{s:realworld-soundness}
\Cref{graph:by_modifications_realworld} categorizes modifications by complexity, and shows cases in which notebook systems' reactions are unsound.

\heading{Direct assignment and reassignment} 
For direct assignment and reassignment modifications---the lower-complexity modifications---both \RRAll and \Ipyflow react correctly. 
This aligns with our expectations, as these modifications rarely involve complex upstream dependencies.
However, \EC fails one direct-assignment case where the modified cell redefines \texttt{data} and \texttt{targets} in terms of their own prior values, requiring an upstream cell to be re-executed first to reset these variables before the modified computation can be reapplied correctly (\S\ref{app:ec-da-failure}).

\heading{Mutation}
As \Cref{graph:by_modifications_realworld}C shows, with more complex modifications, all but \RRAll react incorrectly.
\EC has 3 in-scope mismatches (1 Python built-in mutation and 2 native mutations), for cases that requires upstream reruns to prevent stale state accumulation.
We observe 4 in-scope mismatches for \Ipyflow---2 Python built-in mutations and 2 native mutations---where it misses complex transitive dependencies.
For example, in the \textit{flu\_vaccine} artifact notebook \cite{maasch2025compositional}, the modification inserts a new cell that defines a function to mutate a dataframe in-place, and then immediately calls it:
\begin{verbatim}
def remove_replicate(df, rep_id):
    df.drop(... , inplace=True)
remove_replicate(df_factual, 0)
\end{verbatim}
Since \Ipyflow's tracer relies heavily on symbol resolution within the immediate execution scope, it fails to trace the in-place mutation of \verb|df_factual| through the boundary of the user-defined function. 
When \verb|df_factual| is passed to \verb|remove_replicate|, it is aliased to the local parameter \verb|df|. 
While the \verb|drop| operation mutates the underlying memory object, it is invoked on the local alias \verb|df|. 
\Ipyflow's analyzer fails to perform the deep inter-procedural tracking required to propagate that data-flow dependency back to the global symbol \verb|df_factual|. 
As a result, downstream cells that depend on \verb|df_factual| are not flagged for re-execution, leaving the notebook with an inconsistent state.
Appendix~\ref{app:mutation-categorization} further breaks down the $25$ mutation modifications into sub-categories.

\begin{table}
\caption{\textbf{Realworld In-Scope Rerun Ratios:} Average rerun ratios for all in-scope and matching modifications across three systems. }
\label{tab:realworld_rerun_ratio}
\footnotesize
    \begin{tabular}{rl}
        \toprule
        \textbf{System} & \textbf{Rerun Ratio} \\
        \midrule
        \textbf{\RRAllFS} & 3.30 \\
        \textbf{\EC} & 2.30 \\
        \textbf{\Ipyflow} & 2.80 \\
        \bottomrule
    \end{tabular}
\end{table}

\heading{Out-of-scope cases}
Modifications related to the filesystem or randomness are out-of-scope for \EC and \Ipyflow (5 test cases total).
These systems do not model changes in files or RNG state and thus fail to alert users about potential errors, but sometimes happen to react correctly (4/5 cases for \EC and 1/5 cases for \Ipyflow).



%


\subsection{Precision}
Since rerunning cells can be expensive ~\cite{chattopadhyay2020notebookpain},
we measure precision, or the reactive notebook's computational efficiency with respect to modifications.
We evaluate this using \emph{rerun ratio}, defined as $\frac{|C_{sys}|}{|C_{E}|}$, where $C_{sys}$ is the number of cells re-executed by the system, and $C_{E}$ is the minimal required set (including the modified or added cell).
A rerun ratio of 1 indicates perfect precision and the closer to 1, the better. 
\Cref{tab:realworld_rerun_ratio} shows the average rerun ratios across all in-scope matching benchmarks for different systems.

As expected, \RRAll exhibits the worst precision across all modification types, rerunning 3 to 4 times more cells than necessary (i.e., all cells in the notebook).
\EC generally achieves the lowest rerun ratio since it naively ignores all upstream dependencies.
However, this leads to incorrect states in the more complex cases as discussed in \S\ref{s:realworld-soundness}.
\Ipyflow yields rerun ratios of at least 2. 
This indicates that \Ipyflow's implementation is conservative, commonly triggering cascading reruns of unrelated or independent cells in the same notebook.

\begin{figure}[t!]
\centering
\begin{tikzpicture}
    \begin{axis}[
        ybar,
        bar width=9pt,
        width=6.5cm,
        height=1cm,
        ymin=0,
        ymax=15,
        scale only axis,
        axis x line*=bottom,
        axis y line*=left,
        xtick={1, 2, 3, 4, 5, 6, 7, 8},
        xticklabels={forecasting-gmm, explore\_siren, llm-elicited-priors, lpips, QuEST-mse-fitting, GGAE-public, complexity\_analysis, device\_review},
        x tick label style={
            anchor=north,
            yshift=-10pt,
            xshift=-18pt,
            rotate=45,
            font=\tiny
        },
        xmin=0,
        xmax=9,
        ytick={0, 5, 10, 15},
        yticklabel style = {font=\tiny},
        tick style={draw=none},
        ylabel = {Average Wasted Time (s)},
        ylabel style={font=\footnotesize, yshift=-0.3cm, xshift=0.5cm}, 
        name=bottomaxis
    ]
        \draw[dashed] (axis cs:0,0) -- (axis cs:12.5,0);
        
        \addplot [fill=MyDarkBlue, draw=black] coordinates {
            (1, 1753.6)
            (2, 1501)
            (3, 11.7)
            (4, 11.4)
            (5, 6.3)
            (6, 1.2)
            (7, 1.1)
            (8, 0.2)
        };
    \end{axis}

    \begin{axis}[
        ybar,
        bar width=9pt,
        width=6.5cm,
        height=1.5cm,
        at={(bottomaxis.north west)}, 
        yshift=0.2cm, 
        ymin=1400,
        ymax=1800,
        scale only axis,
        axis x line=none, 
        axis y line*=left, 
        xtick=\empty,
        xmin=0,
        xmax=9,
        ytick={1500, 1800},
        ytick distance=300,
        yticklabel style = {font=\tiny},
        tick style={draw=none},
        name=topaxis
    ]
        \addplot [fill=MyDarkBlue, draw=black] coordinates {
            (1, 1753.6)
            (2, 1501)
            (3, 11.7)
            (4, 11.4)
            (5, 6.3)
            (6, 1.2)
            (7, 1.1)
            (8, 0.2)
        };
    \end{axis}

    \draw[thick] ([xshift=-3pt,yshift=-3pt]topaxis.south west) -- ([xshift=3pt,yshift=3pt]topaxis.south west);
    \draw[thick] ([xshift=-3pt,yshift=-3pt]bottomaxis.north west) -- ([xshift=3pt,yshift=3pt]bottomaxis.north west);

\end{tikzpicture}
\caption{\textbf{Average wasted time (s).} Most wasted time stems from spuriously re-fitting models. We elide a further 4 notebook modifications with negligible wasted time from the graph.}
\label{graph:wasted_time_absolute_broken}
\end{figure}
\heading{Practical impact}
To contextualize \Ipyflow's conservative rerun ratios, we measured the absolute execution time required to execute the extra notebook cells.
While the majority of our benchmarks executed these redundant cells in under 10 seconds due to their relatively lightweight data processing, \Cref{graph:wasted_time_absolute_broken} visualizes the severe penalty deep learning workloads incur when the notebook unnecessarily re-executes cells. 
For example, \Ipyflow introduces over 1,500 seconds of wasted execution time in response to a minor modification to the \textit{explore\_siren} notebook~\cite{sitzmann2020implicit}. 
This notebook contains three computationally expensive but independent model-fitting pipelines: fitting an image, an audio signal, and solving Poisson's equation. 
For example, for a hyperparameter modification to an optimizer's learning rate in one case, a precise system should only re-initialize the model, re-run the image training loop and its subsequent visualization cells. 
Instead, \Ipyflow automatically re-runs unrelated code paths, third-party library imports, class definitions, and the entirely independent audio fitting and Poisson solving pipelines.

\section{Synthetic Notebooks}
\label{s:synthetic}
We next describe results for \RRAll, \EC, \Ipyflow, \Marimo, and \Observable on our synthetic notebooks.
Recall that we design these benchmarks to span a wide variety of modifications.

\begin{figure}[t!]
\centering
\begin{tabular}{c}

\hspace{-0.1cm}
\ref{sharedlegend}
\\[-0.3em]
\hspace{-0.3cm}
\begin{tikzpicture}
    [
      /pgfplots/every axis/.style={
        title = (A) Direct Assignment (18), 
        title style={at={(0.3,0)},anchor=north,yshift=-29pt, font=\footnotesize}, 
        height = 3cm,
        width = 3.3cm,
        symbolic x coords={\RRAllFS, \ECGraphShort, \IpyflowES, \Marimo, \Observable},
        ytick={0, 5, 10, 15, 20}, 
        legend to name=sharedlegend,
        legend style={text=black, xshift=5pt, anchor=center, legend columns=3, font=\footnotesize, nodes={anchor=west}},
        x tick label style={
    anchor=north,                     
    yshift=-8pt, 
    xshift=-13pt, 
    rotate=45,
    font=\tiny
},
        xtick=data,
        ybar stacked,
        ymin=0, 
        ymax=20, 
        ylabel = {Count},
        ylabel style={yshift=-16, xshift=-5, font=\footnotesize}, 
        bar width=7pt,
        enlarge x limits= 0.15, 
        nodes near coords,
        point meta = explicit symbolic,
        ytick distance=10, 
        yticklabel style = {font=\tiny},
        tick style={draw=none}
    }, 
    ]
    \begin{axis}
        \addplot [fill=MyDarkBlue, draw = black, forget plot, font=\tiny, node near coord style={xshift=-13.5pt, yshift=-1.2pt}] coordinates {
        (\RRAllFS,  18)
        (\ECGraphShort, 18)
        (\IpyflowES, 18)
        (\Marimo,  18)
        (\Observable, 8)
    };
    \addplot+[fill = MyPurple, draw = black, forget plot, font=\tiny, node near coord style={xshift=-13.5pt, yshift=-1.2pt}] coordinates {
        (\RRAllFS,   0)
        (\ECGraphShort, 0)
        (\IpyflowES, 0)
        (\Marimo,  0)
        (\Observable, 0)
    };
    \addplot+[fill=MyBeige, draw = black, forget plot, font=\tiny, node near coord style={xshift=-13.5pt, yshift=-1.2pt}] coordinates {
        (\RRAllFS,   0)
        (\ECGraphShort, 0)
        (\IpyflowES, 0)
        (\Marimo,  0)
        (\Observable, 0)
        };
    \addplot+[fill=MyDarkBeige, draw = black, forget plot, font=\tiny, node near coord style={xshift=-13.5pt, yshift=-1.2pt}] coordinates {
        (\RRAllFS,   0)
        (\ECGraphShort, 0)
        (\IpyflowES, 0)
        (\Marimo,  0)
        (\Observable, 0)
    };
    \addplot+[pattern = north east lines, draw = black, forget plot, font=\tiny, node near coord style={xshift=-13.5pt, yshift=-1.2pt}] coordinates {
        (\RRAllFS,   0)
        (\ECGraphShort, 0)
        (\IpyflowES, 0)
        (\Marimo,  0)
        (\Observable, 10)
    };
    
    \addlegendimage{my legend = MyDarkBlue}
    \addlegendentry{In-Scope Match}
    \addlegendimage{my legend = MyPurple}
    \addlegendentry{Out-of-Scope Caught}
    \addlegendimage{legendnortheastlines}
    \addlegendentry{NA}
    \addlegendimage{my legend = MyDarkBeige}
    \addlegendentry{In-Scope Mismatch}
    \addlegendimage{my legend = MyBeige}
    \addlegendentry{Out-of-Scope Not Caught}

    \end{axis}

\end{tikzpicture}

\begin{tikzpicture}
    [
    /pgfplots/every axis/.style={
        title = (B) Reassignment (14), 
        title style={at={(0.35,0)},anchor=north,yshift=-29pt, font=\footnotesize}, 
        height = 3cm,
        width = 3.3cm,
        symbolic x coords={\RRAllFS, \ECGraphShort, \IpyflowES, \Marimo, \Observable},
        ytick={0, 5, 10, 15}, 
        legend style={text=black, xshift=5pt, anchor=center, legend columns=3, font=\footnotesize, nodes={anchor=west}},
        x tick label style={
    anchor=north,                     
    yshift=-8pt, 
    xshift=-13pt, 
    rotate=45,
    font=\tiny, 
    tick style={draw=none}
},
        xtick=data,
        ybar stacked,
        ymin=0, 
        ymax=15, 
        bar width=7pt,
        enlarge x limits= 0.15, 
        nodes near coords,
        point meta = explicit symbolic,
        ytick distance=20, 
        yticklabel style = {font=\tiny},
    }, 
    ]
    \begin{axis}
        \addplot [fill=MyDarkBlue, draw = black, forget plot, font=\tiny, node near coord style={xshift=-13.5pt, yshift=-1.2pt}] coordinates {
        (\RRAllFS,  14)
        (\ECGraphShort, 3)
        (\IpyflowES, 13)
        (\Marimo,  0)
        (\Observable, 0)
    };
    \addplot+[fill = MyDarkBeige, draw = black, forget plot, font=\tiny, node near coord style={xshift=-13.5pt, yshift=-1.2pt}] coordinates {
        (\RRAllFS,   0)
        (\ECGraphShort, 11)
        (\IpyflowES, 1)
        (\Marimo,  0)
        (\Observable, 0)
    };
    \addplot+[fill=MyBeige, draw = black, forget plot, font=\tiny, node near coord style={xshift=-13.5pt, yshift=-1.2pt}] coordinates {
        (\RRAllFS,   0)
        (\ECGraphShort, 0)
        (\IpyflowES, 0)
        (\Marimo,  0)
        (\Observable, 0)
        };
    \addplot+[fill=MyPurple, draw = black, forget plot, font=\tiny, node near coord style={xshift=-13.5pt, yshift=-1.2pt}] coordinates {
        (\RRAllFS,   0)
        (\ECGraphShort, 0)
        (\IpyflowES, 0)
        (\Marimo,  14)
        (\Observable, 14)
    };
    \end{axis}

\end{tikzpicture}

\begin{tikzpicture}
   [
    /pgfplots/every axis/.style={
        title = (C) Mutation (55), 
        title style={at={(0.35,0)},anchor=north,yshift=-29pt, font=\footnotesize}, 
        height = 3cm,
        width = 3.3cm,
        symbolic x coords={\RRAllFS, \ECGraphShort, \IpyflowES, \Marimo, \Observable},
        ytick={0, 20, 40, 60}, 
        legend style={text=black, xshift=5pt, anchor=center, legend columns=3, font=\footnotesize, nodes={anchor=west}},
        x tick label style={
    anchor=north,                     
    yshift=-8pt, 
    xshift=-13pt, 
    rotate=45,
    font=\tiny, 
    tick style={draw=none}
},
        xtick=data,
        ybar stacked,
        ymin=0, 
        ymax=60, 
        bar width=7pt,
        enlarge x limits= 0.15, 
        nodes near coords,
        point meta = explicit symbolic,
        ytick distance=20, 
        yticklabel style = {font=\tiny},
    }, 
    ]
    \begin{axis}
        \addplot [fill=MyDarkBlue, draw = black, forget plot, font=\tiny, node near coord style={xshift=-13.5pt, yshift=-1.2pt}] coordinates {
        (\RRAllFS,  55)
        (\ECGraphShort, 22)
        (\IpyflowES, 34)
        (\Marimo,  0)
        (\Observable, 0)
    };
    \addplot+[fill = MyDarkBeige, draw = black, forget plot, font=\tiny, node near coord style={xshift=-13.5pt, yshift=-1.2pt}] coordinates {
        (\RRAllFS,   0)
        (\ECGraphShort, 23)
        (\IpyflowES, 11)
        (\Marimo,  0)
        (\Observable, 0)
    };
    \addplot+[fill=MyBeige, draw = black, forget plot, font=\tiny, node near coord style={xshift=-13.5pt, yshift=-1.2pt}] coordinates {
        (\RRAllFS,   0)
        (\ECGraphShort, 10)
        (\IpyflowES, 9)
        (\Marimo,  55)
        (\Observable, 14)
    };
    \addplot+[fill=MyPurple, draw = black, forget plot, font=\tiny, node near coord style={xshift=-13.5pt, yshift=-1.2pt}] coordinates {
        (\RRAllFS,   0)
        (\ECGraphShort, 0)
        (\IpyflowES, 0)
        (\Marimo,  0)
        (\Observable, 0)
    };

    \addplot+[pattern=north east lines, draw = black, forget plot, font=\tiny, node near coord style={xshift=-13.5pt, yshift=-1.2pt}] coordinates {
        (\RRAllFS,   0)
        (\ECGraphShort, 0)
        (\IpyflowES, 0)
        (\Marimo,  0)
        (\Observable, 41)
    };

    \addplot+[fill=MyDarkBlue, pattern=north east lines, draw = black, forget plot, font=\tiny, node near coord style={xshift=-13.5pt, yshift=-1.2pt}] coordinates {
        (\RRAllFS,   0)
        (\ECGraphShort, 0)
        (\IpyflowES, 1)
        (\Marimo,  0)
        (\Observable, 0)
    };
    \end{axis} 
\end{tikzpicture}

\end{tabular}
\caption{\textbf{Correctness:} Behavior upon modification for in-scope features (\textit{in-scope match} vs. \textit{in-scope mismatch}) and out-of-scope features (\textit{out-of-scope caught} vs. \textit{out-of-scope not caught}). 51 benchmarks were not comparable (NA) with Javascript-only \Observable due to, e.g., untranslatable Python libraries.}
\label{graph:by_modifications}
\end{figure}

\subsection{Soundness}
While all reactive systems achieved consistent reactivity for direct assignments, reassignment introduces inconsistent behavior; \Ipyflow fails one such case. 
With mutations, systems either fail to detect unsupported  modifications or react inconsistently.
Thus, the data supports our expectation that more complex modifications are harder to handle.

\heading{Direct assignment} 
\textit{Direct assignment} modifications are in-scope for evaluated systems, and all systems correctly identified and re-executed at least the expected set of cells for these benchmarks (Figure \ref{graph:by_modifications}A). 
This consistency suggests direct assignment modifications represent a baseline capability for systems with respect to soundness. 

\heading{Reassignment}
\textit{Reassignment} modifications introduce greater complexity. 
While \Marimo and \Observable place reassignment out of scope (Table~\ref{table:services}), both systems statically detect these modifications and prevent users executing notebooks containing these features ~\cite{marimo_reactivity_guide, larkworthy_observable_notes}. 
This behavior is consistent with their declared limitations and demonstrates that rejecting unhandled executions can be a viable fallback when precise dependency tracking is not available. 
\EC does not provide any handling of cells' upstream dependencies, and thus only matches 3 of the 14 cases.

\Ipyflow, which supports reassignment, correctly handles all but one test case: \texttt{aliasing\_val\_swap}.
In this benchmark, modifying the definition of \texttt{b} requires re-executing the definitions of both \texttt{a} and \texttt{b} as well as their dependents, since a downstream cell performs a simultaneous tuple swap \texttt{a, b = b, a} that makes both variables depend on each other's prior values.
However, \Ipyflow fails to include the definition of \texttt{a} as an upstream dependency, leaving \texttt{b} with a stale value after the swap.
The root cause is that tuple unpacking violates version-ordering assumptions in \Ipyflow's analyzer; we detail this failure and a workaround in Appendix~\ref{appendix:aliasing_val_swap}.


\heading{Mutation}\label{mutation}
No evaluated system with mutations in-scope achieves consistent soundness on our mutation benchmarks other than the baseline, \RRAll (Figure~\ref{graph:by_modifications}C). 
While \RRAll trivially succeeds by (1) over-executing and (2) deleting all created files prior to each modification, all other selective re-execution strategies exhibit significant failure modes. 
\Marimo and \Observable do not support mutation and fail to prohibit or alert users of mutation modifications~\cite{marimo_reactivity_guide, macwright2018observable}. 

\Ipyflow, which attempts to support mutation, has mixed efficacy, correctly handling 34/45 of in-scope test cases.
We defer a detailed walkthrough of these cases to Appendix~\ref{app:mutation_func_boundary}.
Some failures are related to native library extensions: \Ipyflow's dependency tracker operates at the Python symbol level and thus cannot observe mutations performed through compiled library internals, whether via Pandas in-place operations, NumPy C-level functions, scikit-learn training pipelines, or foreign runtime boundaries.



\heading{Out-of-scope cases}
Other than \RRAll, which resets filesystem state before restarting, reactive systems are not aware of interactions with external resources or sources of randomness (10 in total).
Similar to the real-world results, \Ipyflow and \EC do not warn users about these cases (resulting in an Out-Of-Scope Not Caught classification), but sometimes happen to rerun the correct cells.
Specifically, \Ipyflow happens to succeed in one test case where a cell rerun overwrote a file's previous value.
For \Marimo, all test cases result in Out-Of-Scope Not Caught.


\begin{table}[t]
    \centering
    \caption{\textbf{In-Scope Matching Rerun Ratios:} Average rerun ratios of each reactive system on all in-scope, matching benchmarks across modification categories.}
    \label{tab:rerun_ratios}
\footnotesize
    \begin{tabular}{rl}
        \toprule
        \textbf{System} & \textbf{Rerun Ratio} \\
        \midrule
        \RRAll   & 1.56 \\
        \EC         & 1.09 \\
        \Ipyflow    & 1.24 \\
        \Marimo     & 1.00 \\
        \Observable & 1.00 \\
        \bottomrule
    \end{tabular}
\end{table}

\subsection{Precision}
Table~\ref{tab:rerun_ratios} reports the rerun ratio, as defined in \S\ref{s:realworld}, for in-scope test cases for each notebook.
\Marimo and \Observable achieve the ideal rerun ratio of 1, as they limit the scope of allowed programs and modifications.
The limited scope allows \Observable and \Marimo to track dependencies and the minimal set of cells to rerun, with only static analysis techniques~\cite{agrawal2024lessons, macwright2018observable}.
\RRAll, \EC, and \Ipyflow follow a similar trend as the real-world results, for similar reasoning.
\RRAll has the highest rerun ratio, followed by \Ipyflow, followed by \EC.



\section{Discussion and Related Work}
\label{s:5}

So, when are reactive notebooks not reactive? \Marimo and \Observable clearly are not reactive for programs out of scope, suggesting the need for better analysis tools which can detect undefined behavior and alert users.
\Ipyflow fails to react to a handful of edge cases, although there are few clear patterns to the failures. All the systems we tested failed to react to changes to the external environment, as expected.

Our results demonstrate the need for reactive notebook systems to more clearly communicate their capabilities and limitations to their users, through both documentation and tooling.
These systems do not need to be perfect, but users at least need a reasonable model of when they can expect a given system to function correctly. We hope that benchmarks like \sys will provide one avenue for system designers to identify system limitations.
For example, we note that spreadsheets, another set of traditional data analysis tools, provide ironclad reactivity so that users can always be confident in the effect their updates will have.

Additionally, we expect that \sys will spawn avenues of research that address the gaps classified in existing systems, as prior benchmarks have for, \eg the Unix shell~\cite{koalas}.
For example, we imagine that reactivity systems can draw on system tracing, sandboxing, and checkpointing techniques to monitor interactions with external system state and correctly react to modifications involving external state.
Side-effect control of arbitrary components in combination with just-in-time multi-grained dependency reconstruction can help track and respond to changes at levels appropriate for the granularity of changes performed by their users.
Orthogonally, ways to annotate library interfaces can help reactivity systems understand how external library functions read and mutate program state.
Finally, research on state tracking in language-agnostic intermediate representations by leveraging, for example, web assembly, can help create reactivity systems capable of working in any programming language.

\heading{Beyond Reactivity}
Reactivity is not the only solution for improving state management and user experience with computational notebooks. An alternative solution is taking snapshots of the current notebook state for users to rollback or maintain more organized exploration process through version control~\cite{10.1145/3411764.3445527, 10.14778/3626292.3626296, 10.14778/3717755.3717759, sato2024multiverse, 10.14778/3514061.3514075, 8506576}. 
Similarly, users may choose to traverse multiple isolated exploration paths by forking~\cite{fang2025enhancing}. 
Lineage tracking and annotations in notebooks further assist users in associating related cells and quickly identify cells needed to produce certain outputs~\cite{nbgather, 10.1145/3290605.3300500, flowbook, 10.1145/3274419}, increasing reproducibility of notebooks. 
\sys can be adapted for continued development of these solutions or future works.

\section{Conclusion}
Computational notebooks are interesting because they provide an execution model that allows users to modify programs as they run them. This makes notebooks a natural fit for writing exploratory programs. However, it is common for notebook state to become messy and unreproducible. Reactive notebooks are a popular solution for handling state management on-behalf of the users. Different from other reactive applications such as spreadsheets, reactive notebooks flip the feature set: they support arbitrary programs and thus provide users more flexibility, but as we have seen, offer inconsistent and unpredictable reactivity implementations instead.
The success of spreadsheet applications shows that users benefit from reactivity when it provides consistent results.
We thus call for research on bringing this ironclad reactivity to computational notebooks, which will unlock an entire class of exploratory programming held back by inconsistent implementations today.

\clearpage
\bibliographystyle{ACM-Reference-Format}
\bibliography{sample-base}

\appendix

\setcounter{section}{0}
\renewcommand{\thesection}{\Alph{section}}

\setcounter{table}{0}
\renewcommand{\thetable}{\Alph{section}\arabic{table}}

\setcounter{figure}{0}
\renewcommand{\thefigure}{\Alph{section}\arabic{figure}}

\clearpage

\section{Mutation Categorization}
\label{app:mutation-categorization}

\begin{table*}[t]
\caption{\textbf{Mutation subcategories in \bench:} definitions, examples, and IPyflow failure counts across realworld (RW) and synthetic (Syn) benchmarks.}
\label{tab:mutation-subcategories}
\small
\begin{tabularx}{\textwidth}{lXXlllll}
\toprule
\textbf{Subcategory} & \textbf{Definition} & \textbf{Examples} & \textbf{RW} & \textbf{RW Fail} & \textbf{Syn} & \textbf{Syn Fail} \\
\midrule
Python built-in mutations &
    In-place modifications via pure Python semantics: container methods, subscript/key assignment, augmented assignment on mutables, aliasing, and mutation through user-defined function/method boundaries. &
    \texttt{b = a; b.append(1)}, \texttt{x[0] = 6}, \texttt{lst.append(x)} &
    11 & 2 & 31 & 9 \\
\midrule
Native mutations &
    In-place modifications via native/compiled library code (numpy, pandas) where mutation semantics (views, \texttt{inplace=True}, in-place library functions) require library-specific knowledge. &
    \texttt{DataFrame.drop(..., inplace=True)} &
    9 & 2 & 14 & 2 \\
\midrule
External state &
    Modifications to state outside the Python object graph: file system, foreign runtimes, databases, or library global singletons. Note that these are out-of-scope for IPyFlow, so failure indicates not alerting the user.&
    \texttt{open()}+write, \texttt{np.save()}+read, 
    &
    5 & 4 & 10 & 9 \\
\bottomrule
\end{tabularx}
\end{table*}

Table~\ref{tab:mutation-subcategories} breaks down the three mutation subcategories by benchmark count and failure rate across real-world and synthetic benchmarks.
Python built-in mutations are the most prevalent subcategory, with 12 real-world and 31 synthetic cases.
\Ipyflow handles the majority correctly---failing on 2 real-world and 9 synthetic cases, primarily due to function boundary and nested collection patterns described in Appendix \ref{app:mutation_func_boundary}.
For native mutations, \Ipyflow fails on 2 of 9 real-world and 2 of 14 synthetic cases, as its symbol-level tracker cannot observe mutations performed through compiled library internals.
External state mutations are the most problematic subcategory. 
\Ipyflow fails on 4/5 real-world and 9/10 synthetic cases, consistent with its declared out-of-scope treatment of filesystem and foreign runtime interactions.
\Ipyflow fails as it does not indicate to the user that such notebook modifications are out-of-scope, but happens to react correctly in 1/5 and 1/10 scenarios, respectively.
Together, these results show that failure rates increase with the degree to which mutation semantics are hidden from Python's symbol-level dependency tracking.

%

\section{Modification Generation}
\label{app:prompt}

We used Claude~4.6~Sonnet via Claude Code to automate the initial modification generation across all \numRealworldNotebook notebooks. For each notebook, we issued the following prompt:

\begin{quote}
\ttfamily\small\raggedright
Read the ipynb file and come up with 30 realistic modifications that a common developer would make either to achieve this state or to modify based on this state. Here is the modification requirement:
\begin{itemize}
    \item It must be realistic and make sense.
    \item Modifying this cell would lead to implicit/explicit changes in other cells that use data in the modified cell.
    \item The modification needs to be applied to data that will flow into a C-extension. Bonus if the C-extension implicitly modifies the data passed to it. For example, cell 1 calls \texttt{cfunc(A)} where \texttt{cfunc} reads and writes to \texttt{A} but cannot be detected by reactive notebooks like \Ipyflow.
    \item It can also involve I/O. For example, \texttt{torch.save(x)} in cell 2 and \texttt{torch.load(x)} in cell 3, with a modification to cell 1 where \texttt{x} is defined. We expect cell 3 to be rerun, but reactive notebooks like \Ipyflow cannot handle this.
\end{itemize}
\end{quote}
All proposed modifications were manually reviewed and corrected. 
We verified executability, confirmed the minimal expected rerun set by reasoning through each notebook's execution topology, and empirically validated each set against a complete top-to-bottom re-execution from a fresh environment, yielding the final \rNum modifications used in evaluation.

\section{Mutation Example}
\label{app:mutation_example}


\begin{figure}[t!]
    \centering

\includegraphics[scale=0.41]{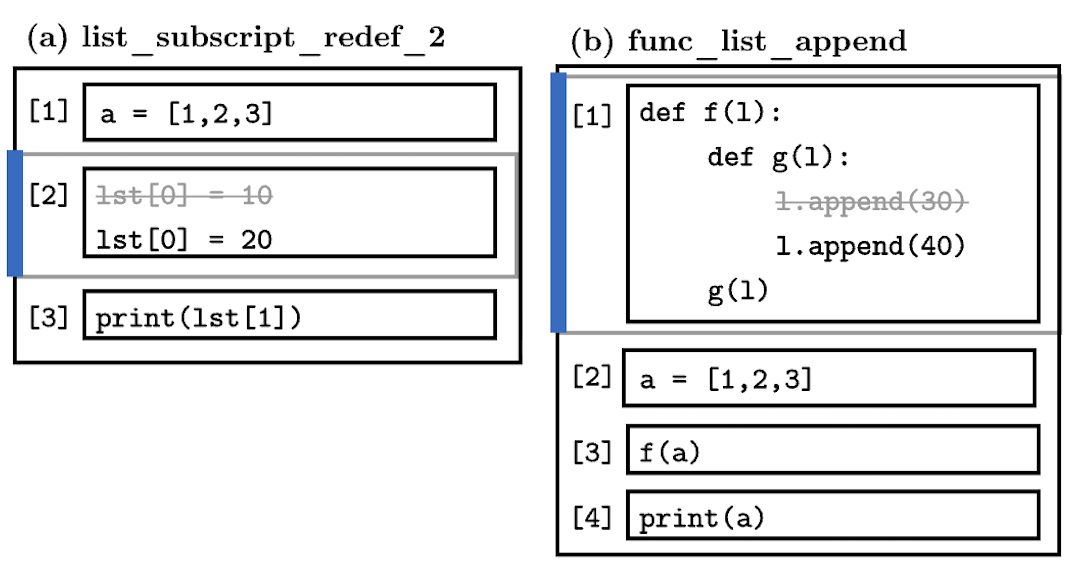}


\caption{\textbf{Collection And Function Mutations:} Mutation benchmarks containing collections or through functions can be challenging for reactive systems to balance consistency with efficiency. Cell 2 in \texttt{list\_subscript\_redef\_2} and cell 1 in \texttt{func\_list\_append} above contains the modification, depicted with the original code crossed-out and replaced by the user modification.} 
\label{fig:ipyflow-mutations}
\end{figure}

\subsection{IPyflow Mutation Failures: Collection and Function Boundary Cases}
\label{app:mutation_func_boundary}

\heading{Collection mutations}
IPyflow correctly handles mutations that are idempotent on the referenced object and whose downstream cells access unaffected elements.
For example, in \texttt{list\_subscript\_redef\_2} (Figure~\ref{fig:ipyflow-mutations}a), the modification changes \texttt{lst[0]}, but the only downstream cell reads \texttt{lst[1]}, so only the mutated cell need be re-executed, which \Ipyflow correctly identifies.

However, \Ipyflow fails when a mutation to a nested key or index is transitively read downstream.
In Figure~\ref{fig:intro-fig}, the value \texttt{x['a']} is mutated in one cell, but \Ipyflow fails to detect that a later \texttt{print(x)} transitively depends on key \texttt{'a'}.
Consequently, the original definition of \texttt{x} is not re-executed, and the notebook prints a stale value.
After we contacted the \Ipyflow developers about this behavior, they issued several fixes to their nested collection tracking~\cite{ipyflow}.
These cases illustrate that even when the root cause is an implementation bug rather than a fundamental limitation, the sheer variety of mutation patterns makes exhaustive correctness difficult to achieve.

\heading{Function boundary mutations}
2 of the 11 failures in synthetic notebook benchmarks occur when a mutation is performed inside a user-defined function that receives a mutable object as an argument.
In \texttt{func\_list\_append} (Figure~\ref{fig:ipyflow-mutations}b), the nested function \texttt{g(l)} appends to its argument \texttt{l} in place.
When \texttt{f(a)} is called, \texttt{a} is aliased to \texttt{l} inside the function; the append mutates the underlying object, but \Ipyflow only observes the call site symbol \texttt{a} and does not perform inter-procedural tracking to propagate the side effect back to downstream cells that read \texttt{a}.
As a result, \texttt{print(a)} is not flagged for re-execution and produces a stale value.
This pattern---mutation through a pass-by-reference function argument---is the same subcategory responsible for IPyflow's \texttt{flu\_vaccine} failure in real-world notebooks (\S\ref{s:realworld}).

\subsection{\EC Failure Case: Self-Referential Redefinition}
\label{app:ec-da-failure}

In the \textit{GGAE-public}~\cite{lim2024graph} notebook, cell 2 loads the dataset's raw tensors:
\begin{verbatim}
d = dl.dataset.data
t = dl.dataset.targets
\end{verbatim}
Cell 3 then defines a constant \texttt{EPI\_PER\_IMAGE} and uses it to expand \texttt{data} and \texttt{targets} in place, redefining each variable in terms of its own previous value:
\begin{verbatim}
EPI_PER_IMAGE = 1
d = d.repeat_interleave(EPI_PER_IMAGE, 0)
t = t.repeat_interleave(EPI_PER_IMAGE, 0)
\end{verbatim}
The modification changes \texttt{EPI\_PER\_IMAGE} from \texttt{1} to \texttt{3}, a direct assignment within cell 3.
\EC re-executes cell 3 and all spatially subsequent cells, but does not re-execute cell 2.
Since cell 3's computation of \texttt{d} and \texttt{t} reads and overwrites the \emph{same} variables it depends on, re-running cell 3 alone applies \texttt{repeat\_interleave} a second time to the \emph{already-expanded} tensors from the first execution.
A sound reaction requires re-executing cell 2 first, to reset \texttt{d} and \texttt{t} to their original, un-expanded values, before cell 3 is re-executed with the updated constant.

\subsection{IPyflow Failure: \texttt{aliasing\_val\_swap}}
\label{appendix:aliasing_val_swap}

\begin{figure}[t!]
    \centering
\begin{tikzpicture}[x=0.90pt,y=0.78pt,yscale=-1,xscale=1]

\draw  [draw opacity=0][fill={rgb, 255:red, 251; green, 247; blue, 197 }  ,fill opacity=1 ] (76,60) -- (176.5,60) -- (176.5,73.5) -- (76,73.5) -- cycle ;
\draw  [line width = 0.75] [color={rgb, 255:red, 155; green, 155; blue, 155 }  ,draw opacity=1 ] (58.75,57.63) -- (181.5,57.63) -- (181.5,78) -- (58.75,78) -- cycle ;
\draw  [line width=0.75]  (76,41) -- (176.5,41) -- (176.5,54.5) -- (76,54.5) -- cycle ;
\draw  [line width = 0.75] (76,81) -- (176.5,81) -- (176.5,92.5) -- (76,92.5) -- cycle ;
\draw   [line width = 0.75] (58.75,35.25) -- (181.5,35.25) -- (181.5,115) -- (58.75,115) -- cycle ;
\draw [color={rgb, 255:red, 34; green, 109; blue, 196 }  ,draw opacity=1 ][line width=3]    (58.75,57.63) -- (58.75,78) ;
\draw  [line width=0.75]  (76,60) -- (176.5,60) -- (176.5,73.5) -- (76,73.5) -- cycle ;
\draw   [line width = 0.75] (76,97) -- (176.5,97) -- (176.5,108.5) -- (76,108.5) -- cycle ;
\draw  [line width = 0.75] [draw opacity=0][fill={rgb, 255:red, 251; green, 247; blue, 197 }  ,fill opacity=1 ] (211,60) -- (311.5,60) -- (311.5,73.5) -- (211,73.5) -- cycle ;
\draw [line width = 0.75]  [color={rgb, 255:red, 155; green, 155; blue, 155 }  ,draw opacity=1 ] (193.75,57.63) -- (316.5,57.63) -- (316.5,78) -- (193.75,78) -- cycle ;
\draw  [line width=0.75]  (211,41) -- (311.5,41) -- (311.5,54.5) -- (211,54.5) -- cycle ;
\draw   [line width = 0.75] (211,81) -- (311.5,81) -- (311.5,92.5) -- (211,92.5) -- cycle ;
\draw   [line width = 0.75] (193.75,35.25) -- (316.5,35.25) -- (316.5,115) -- (193.75,115) -- cycle ;
\draw [color={rgb, 255:red, 34; green, 109; blue, 196 }  ,draw opacity=1 ][line width=3]    (193.75,57.63) -- (193.75,78) ;
\draw  [line width=0.75]  (211,60) -- (311.5,60) -- (311.5,73.5) -- (211,73.5) -- cycle ;
\draw   [line width = 0.75] (211,97) -- (311.5,97) -- (311.5,108.5) -- (211,108.5) -- cycle ;

\draw (78,44) node [anchor=north west][inner sep=0.75pt]  [font=\tiny] [align=left] {{\fontfamily{pcr}\selectfont \textbf{a = 9}}};
\draw (74,44) node [anchor=north east] [inner sep=0.75pt]  [font=\tiny] [align=left] {{\fontfamily{pcr}\selectfont \textbf{[1]}}};
\draw (79,64) node [anchor=north west][inner sep=0.75pt]  [font=\tiny] [align=left] {{\fontfamily{pcr}\selectfont \textbf{b = 5}}};
\draw (74,64) node [anchor=north east] [inner sep=0.75pt]  [font=\tiny] [align=left] {{\fontfamily{pcr}\selectfont \textbf{[2]}}};
\draw (78,84) node [anchor=north west][inner sep=0.75pt]  [font=\tiny] [align=left] {{\fontfamily{pcr}\selectfont \textbf{a,b = b,a}}};
\draw (74,84) node [anchor=north east] [inner sep=0.75pt]  [font=\tiny] [align=left] {{\fontfamily{pcr}\selectfont \textbf{[3]}}};
\draw (60.75,32.25) node [anchor=south west] [inner sep=0.75pt]  [font=\tiny] [align=left] {{\fontfamily{pcr}\selectfont \textbf{Original Notebook}}};
\draw (195.75,32.25) node [anchor=south west] [inner sep=0.75pt]  [font=\tiny] [align=left] {{\fontfamily{pcr}\selectfont \textbf{Modified Notebook}}};
\draw (78,100) node [anchor=north west][inner sep=0.75pt]  [font=\tiny] [align=left] {{\fontfamily{pcr}\selectfont \textbf{print("a:", a, "b:", b)}}};
\draw (74,100) node [anchor=north east] [inner sep=0.75pt]  [font=\tiny] [align=left] {{\fontfamily{pcr}\selectfont \textbf{[4]}}};
\draw (213,44) node [anchor=north west][inner sep=0.75pt]  [font=\tiny] [align=left] {{\fontfamily{pcr}\selectfont \textbf{a = 9}}};
\draw (209,44) node [anchor=north east] [inner sep=0.75pt]  [font=\tiny] [align=left] {{\fontfamily{pcr}\selectfont \textbf{[1]}}};
\draw (214,64) node [anchor=north west][inner sep=0.75pt]  [font=\tiny] [align=left] {{\fontfamily{pcr}\selectfont \textbf{b = 8}}};
\draw (209,64) node [anchor=north east] [inner sep=0.75pt]  [font=\tiny] [align=left] {{\fontfamily{pcr}\selectfont \textbf{[2]}}};
\draw (213,84) node [anchor=north west][inner sep=0.75pt]  [font=\tiny] [align=left] {{\fontfamily{pcr}\selectfont \textbf{a,b = b,a}}};
\draw (209,84) node [anchor=north east] [inner sep=0.75pt]  [font=\tiny] [align=left] {{\fontfamily{pcr}\selectfont \textbf{[3]}}};
\draw (213,100) node [anchor=north west][inner sep=0.75pt]  [font=\tiny] [align=left] {{\fontfamily{pcr}\selectfont \textbf{print("a:", a, "b:", b)}}};
\draw (209,100) node [anchor=north east] [inner sep=0.75pt]  [font=\tiny] [align=left] {{\fontfamily{pcr}\selectfont \textbf{[4]}}};

\end{tikzpicture}
\caption{\textbf{\texttt{aliasing\_val\_swap} Notebook Benchmark:} A benchmark contains an \textit{original notebook} and a \textit{modified notebook}. The modification in this example is highlighted in \hl{yellow}.}
\label{figure:aliasing_val_swap}
\end{figure}

Figure~\ref{figure:aliasing_val_swap} shows the \texttt{aliasing\_val\_swap} benchmark.
The one-line tuple swap \texttt{a, b = b, a} in cell 3 updates both variables simultaneously.
The expected reaction to modifying cell 2 (\texttt{b = 5} $\to$ \texttt{b = 8}) is to re-execute cells 1 and 2 (the definitions of \texttt{a} and \texttt{b}) followed by their dependents, cells 3 and 4.
However, \Ipyflow fails to include cell 1 as an upstream dependent of cell 2, so \texttt{b} is assigned the \emph{original} value of \texttt{a} (i.e., 5) instead of the correct value 9.

Swap operations are particularly challenging because they introduce mutual dependencies between variables, requiring the system to reason not only about data-flow but about the \emph{versioning} of each variable.
While \Ipyflow associates variables with version metadata~\cite{ipyflow}, the simultaneous update of tuple unpacking appears to violate assumptions in its analyzer implementation.
Notably, rewriting the swap with a temporary variable (e.g., \texttt{tmp = a; a = b; b = tmp}) causes \Ipyflow to react correctly, confirming the issue is specific to simultaneous tuple unpacking rather than swap semantics in general.

\end{document}